\documentclass[useAMS,usenatbib,usegraphicx]{mn2e}

\newcommand{\kev}{keV}
\newcommand{\fe}{Fe~K$\alpha$}
\newcommand{\etal}{et al.}
\newcommand{\mcg}{MCG--6-30-15}
\newcommand{\h}{1H~0707--495}

\title[Reflection dominated X-ray spectra in Seyfert 1 galaxies]
  {How the X-ray spectrum of a Narrow-Line Seyfert 1 galaxy may be 
  reflection dominated}
\author[A.\ C.\ Fabian \etal]
  {A.~C.~Fabian$^1$\thanks{acf@ast.cam.ac.uk}, D.~R.~Ballantyne$^1$,
  A.~Merloni$^1$, S.~Vaughan$^1$, K.~Iwasawa$^1$ and Th.~Boller$^2$\\
  $^1$Institute of Astronomy, Madingley Road, Cambridge CB3 0HA \\
  $^2$Max-Planck-Institut f\"{u}r extraterrestrische Physik, Postfach
  1603, 85748 Garching, Germany}

\pagerange{\pageref{firstpage}--\pageref{lastpage}}
\pubyear{2002}

\usepackage{times}

\begin{document}

\label{firstpage}

\maketitle

\begin{abstract}
A model for the inner regions of accretion flows is presented where,
due to disc instabilities, cold and dense material is clumped into
deep sheets or rings. Surrounding these density enhancements is hot,
tenuous gas where coronal dissipation processes occur. We expect this
situation to be most relevant when the accretion rate is close to
Eddington and the disc is radiation-pressure dominated, and so may
apply to Narrow-Line Seyfert~1 (NLS1) galaxies. In this scenario, the
hard X-ray source is obscured for most observers, and so the detected
X-ray emission would be dominated by reflection off the walls of the
sheets. A simple Comptonization calculation shows that the large
photon-indices characteristic of NLS1s would be a natural outcome of
two reprocessors closely surrounding the hard X-ray source. We test
this model by fitting the \textit{XMM-Newton} spectrum of the NLS1 \h\
between 0.5 and 11~\kev\ with reflection dominated ionized disc
models. A very good fit is found with three different reflectors each
subject to the same $\Gamma=2.35$ power-law. An iron overabundance is
still required to fit the sharp drop in the spectrum at around
7~\kev. We note that even a small corrugation of the accretion disc
may result in $\Gamma > 2$ and a strong reflection component in the
observed spectrum. Therefore, this model may also explain the strength
and the variability characteristics of the \mcg\ \fe\ line.  The idea
needs to be tested with further broadband \textit{XMM-Newton}
observations of NLS1s.
\end{abstract}


\section{Introduction}
\label{sect:intro}

The basic model for the active nucleus of a typical Seyfert 1 galaxy
consists of a black hole surrounded by an ultraviolet and soft X-ray
emitting thin accretion disc above which a patchy active corona emits
hard X-rays. The model is supported by strong X-ray reflection
signatures due to Compton backscattering and fluorescence of the
harder coronal X-rays by the dense disk \citep{pou90,nan94,pet00}. In
some objects an extreme red wing is found in the iron line of the
reflection spectrum \citep{tan95, nan99, fab00}, showing that the disk
can extend very close to the black hole where large relativistic
effects occur. In general the strength of the reflection spectrum
indicates that the disk surface is approximately flat, although disk
instabilities could well make it ribbed or clumpy, particularly if the
disk is radiation-pressure dominated
\citep[e.g.,][]{le74,gr88,kro98,tss01}. Such a situation is most
likely to occur when the accretion rate is high and close to the
Eddington limit, which is often thought to explain the unusual
properties of Narrow-Line Seyfert 1 (NLS1) galaxies
\citep*[e.g.,][]{pou95}.

Recently, \citet{bol01} reported on a ``sharp spectral feature'' at
about 7.1~\kev\ in the \textit{XMM-Newton} spectrum of the NLS1 \h\
($z=0.0411$). This deep drop (over a factor of two in flux) in the
spectrum occurs at almost the exact energy of the neutral iron edge,
so absorption models (in particular, a partial covering model) were
favoured to explain the feature. However, these models suffered from a
very soft intrinsic power-law ($\Gamma \sim 3.5$) and an unreasonably
large value of the iron abundance ($\sim$ 35$\times$ solar). Moreover,
it was difficult for the partial covering model to account for the
absence of a narrow \fe\ line and to explain the rapid variability of
the source over the entire waveband. The sharp drop in the spectrum
could also be due to the blue wing of a relativistic iron emission
line, but, as the authors noted, to explain the depth of the drop at
$\sim$7~\kev\ requires ``invoking a very extreme Fe abundance and/or
reflection fraction.'' Here, we consider the possibility of a
reflection dominated X-ray spectrum in detail.

\begin{figure}
\begin{center}
\includegraphics[width=0.35\textwidth,angle=-90]{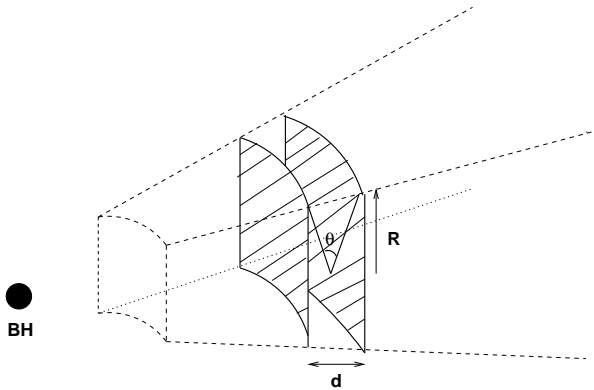}
\caption{The proposed geometry. The inner region of the
radiation-pressure dominated accretion disc is broken up into a series
of deep sheets or rings, with a typical size, $R$, of the order of the disc thickness, and separated by a distance $d$. 
Hard X-ray sources are found in between the
sheets, but are obscured from the view of observers along most
line-of-sights. Therefore, the detected X-ray emission is
dominated solely by the reflected emission from the walls of the sheets. 
In Section~\ref{sect:cand}, we have considered circular sheets of
radius $R$ instead of squared ones to simplify the computations.}
\label{fig:geom1}
\end{center}
\end{figure}

The basic idea we wish to explore is that within a system which is
accreting at close to the Eddington rate, disc instabilities funnel
denser material into many deep rings, between which the hard X-rays
are emitted by some coronal process (Figure~\ref{fig:geom1}). To an
off-axis observer, the X-ray emission will consist mostly of a
reflection component.  The rapid variability seen in the X-ray
emission from \h\ could then in part be due to changes in the geometric
obscuration of one sheet on another, explaining why little spectral
variability is seen \citep{bol01}. This reflection model requires a
less extreme input spectrum and hidden flux than the partial-covering
model.
While this is clearly an idealized picture, recent numerical simulations
of the magneto-rotational instability in radiation-pressure dominated
discs have found that the turbulence can cause density contrasts up to
a factor of 200 \citep{tss01}. Therefore, it does seem plausible that
a radiation-pressure dominated disc will exhibit strong density inhomogeneities.

In the next section, we calculate what the hard X-ray power-law may
look like in our idealized picture and show that a steep spectrum
naturally arises from this situation. In Section~\ref{sect:1h} we
fit reflection dominated spectra to the broadband \textit{XMM-Newton}
data of \h. Finally, we discuss the implications of this model
in Section~\ref{sect:discuss}.
 
\section{Calculation of the X-ray power-law}
\label{sect:cand}

Here we give a simple analytic estimate of the slope of the X-ray
continuum by assuming that in the inner radiation-pressure dominated
part of the accretion disc the dense and cold sheet-like part of the
flow is embedded in a tenuous, hot plasma
\citep[cf.,][]{gr88,cfr92,kbr96,kcr97}. The accretion energy is
dissipated, probably via magnetic reconnection, in the hot phase only,
with the cold sheets reprocessing (reflecting and thermalizing) the
incident hard flux.

For ease of computation, the picture in Fig.~\ref{fig:geom1} was
adapted so that the cold sheets have a flat disc-like geometry with
radius $R$. We consider two such discs facing each other and separated
by a distance $d$. The hot phase is sandwiched by the two cold discs
and therefore has a cylindrical slab geometry. From a point in the
mid-plane of the cylinder, at a distance $r$ from its centre, the
covering fraction of the two cold sheets is
\begin{equation}
C(r)=\left[1-\frac{1}{2}\left(\frac{1}{\sqrt{1+\frac{4(R-r)^2}{d^2}}}+
\frac{1}{\sqrt{1+\frac{4(R+r)^2}{d^2}}}\right)\right].
\end{equation}
The average total covering fraction is then given by
\begin{equation}
C=\frac{1}{R}\int_0^R C(r) dr=1-\frac{d}{4R}\left[\log \left(\frac{4R}{d}+\sqrt{1+\frac{16 R^2}{d^2}}\right)\right].
\end{equation}

With the hard luminosity of the hot phase being $L_{\rm
H}$, the reprocessed luminosity is given by
\begin{equation}
L_{\rm rep}=C L_{\rm H} (1-a),
\end{equation}
where $a$ is the albedo of the cold sheets. We make use of the results
of \citet*{mbp01} who used Monte Carlo simulations of Comptonizing
coronae above dense cold material and showed that the albedo depends
on the spectral index of the illuminating radiation, being smaller for
steep spectra. The thermal reprocessed soft photon flux from the cold
sheets is assumed to be a black-body spectrum with temperature
$T_{\mathrm{rep}}$, and is calculated by the method outlined by
\citet{mf01}.  The hard luminosity is assumed to be due only to
inverse Compton scattered photons and can be approximated as a sum of
a cut-off power-law and Wien component \citep{wz00}.

The emerging spectral index of the power-law is given by
\begin{equation}
\Gamma-1=\alpha=-\frac{\ln P_{\rm sc}}{\ln (1+4\Theta+16\Theta^2)},
\end{equation}
where $\Theta=kT_{\mathrm{e}}/m_{\mathrm{e}}c^2$ 
is the electron temperature of the hot phase, and
$P_{\rm sc}$ is the scattering probability averaged over the source
volume and depends only on the coronal optical depth $\tau$.
In a slab geometry $P_{\rm sc}$ can be approximated as \citep{z94}
\begin{equation}
P_{\rm sc}=1+\frac{\exp(-\tau)}{2}\left(\frac{1}{\tau}-1\right)-\frac{1}{2\tau}+\frac{\tau}{2}E_1(\tau),
\end{equation}
where $E_1$ is the exponential integral.

\begin{figure}
\includegraphics[width=0.3\textwidth,angle=-90]{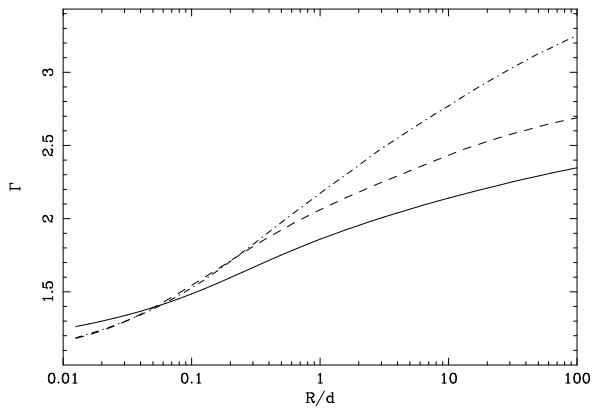}
\caption{The spectral index $\Gamma$ as a function of the aspect ratio
$R/d$, for $\tau=0.1$ (solid line), $\tau=1$ (dashed line) and 
$\tau=10$ (dot-dashed line).} 
\label{fig:gamma_rd}
\end{figure}

Once the geometry is fixed (by fixing the aspect ratio $R/d$), the 
temperature of the hot phase, and consequently the slope of the
Comptonized continuum, can be calculated self-consistently by solving 
$L_{\rm H}=L_{\rm C}=\int^\infty_{2.78\Theta_{\rm rep}}L_{\rm C}(x) dx$.
We have calculated the spectral index $\Gamma$ for different values of the 
ratio $R/d$ and in Figure~\ref{fig:gamma_rd} we plot the spectral
index for three different values of the optical depth of the hot phase.
Clearly, as the covering fraction increases, the spectrum softens 
significantly. Also, the spectrum is steeper for a denser hot medium 
(larger $\tau$). Finally, if there is accretion energy dissipated in
the cold phase, then the spectrum will be softer for a given $R/d$
\citep[c.f.,][]{mal01}.

\begin{figure}
\includegraphics[width=0.3\textwidth,angle=-90]{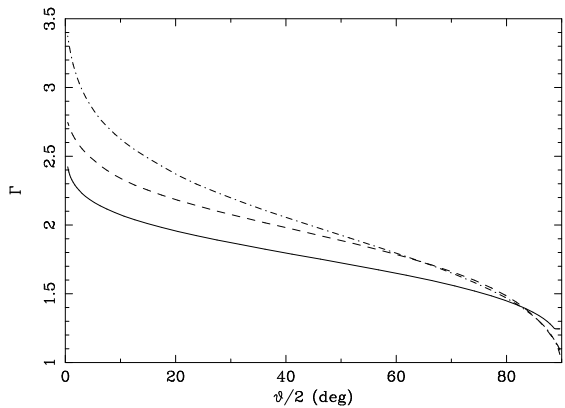}
\caption{The spectral index $\Gamma$ as a function of the
half opening angle of the couple of disc-like cold sheets for $\tau=0.1$ (solid line), $\tau=1$ (dashed line) and $\tau=10$ (dot-dashed line).} 
\label{fig:gamma-th}
\end{figure}

Figure~\ref{fig:gamma-th} shows an equivalent plot where the abscissa
is now the half opening angle between the two reprocessors $\theta/2$.
These plots show that spectral indices typical of NLS1s (i.e., $\Gamma
>$ 2.1--2.2) are a natural consequence of this scenario if the half opening
angle between the two reprocessors is less than about 20 degrees. In
such a case, it is likely that a observer may only detect the
reflected emission arising from the walls of the reprocessors.

\section{Application to the NLS1 1H~0707--495}
\label{sect:1h}

In this section, we attempt to fit the broadband X-ray spectrum of the
NLS1 \h\ with models of reflection dominated spectra, which might be
expected if the accretion disc has broken into a number of dense
segments (Fig.~\ref{fig:geom1}). We consider the EPIC-pn spectrum of
\h\ between 0.5 and 11~\kev\ and fit with the constant density ionized
disc models of \citet{ros93} (see also \citealt*{ros99}). The
parameters of the model are the ionization parameter $\xi=4 \pi
F_{\mathrm{X}}/n_{\mathrm{H}}$, where $F_{\mathrm{X}}$ is the
illuminating flux between 10~eV and 100~\kev, the photon index of the
power-law spectrum that strikes the slab, and a normalization
constant. Relativistic blurring was applied to the model during
fitting using the kernel of \citet{lao91}. The emissivity index was
fixed at $-3$, but the inner and outer radii, along with the
inclination angle, were left free. Absorption due to the Galactic
column of $5.79\times10^{20}$~cm$^{-2}$ was included in all fits.

If one tries to fit the data with a single reflection spectrum, then
the best fit ($\chi^2$/d.o.f.=506/309) is obtained with a model with a
three times solar Fe abundance. Recall that no power-law component has
been added to these models, giving them, in effect, an infinite
reflection fraction. We find that $\Gamma=2.44$ and $\log \xi=2.040$,
which results in a strong neutral \fe\ line. Extreme Kerr blurring
($r_{\mathrm{in}}=1.24$~$r_g$) was needed to fit the large red wing of
the line between 2.5 and 6~\kev. However, as indicated by the large
$\chi^2$, there were significant residuals to the fit, particularly
below 1~\kev\ where the model predicts strong emission lines from Fe
and O \citep{ros93}. To dilute the effect of these lines, a power-law
was added to the model with the value of $\Gamma$ fixed to be the same
as the reflected component. Reflection still dominated the spectrum
with a reflection fraction $>$ 10. The power-law improved the best fit to
$\chi^2$/d.o.f.=479/308, but required an Fe abundance greater than 5
times solar. In this fit, $\Gamma$ softened slightly to 2.52, but
there was little change in the ionization parameter, the inner and
outer radii, or the inclination angle ($\sim 23$~degrees). There are
still residuals below 1~\kev, but, interestingly, there is now a clear
line-like residual just above 6~\kev. Adding a Gaussian to the fit
drops the $\chi^2$ by 38 with the addition of three degrees of freedom
-- significant at $> 99.99$~per cent, according to the F-test. Here,
the best-fit model has a seven times overabundance of Fe, and
$\Gamma=2.56$. The line had an energy of 6.62~\kev\ and a width
$\sigma=0.235$~\kev, indicating that it probably also arises from
somewhere in or on the disk.

To investigate this further, the power-law and Gaussian components of
the model were replace by a second reflector. The second reflection
component was drawn from the same group of models as the first, and
was also subject to relativistic blurring. We fixed the photon-index
and inclination angle to be the same for both components. The addition
of this second reflector greatly improved the fit
($\chi^2$/d.o.f.=342/305), as the combined emission removed most of
the residuals from the soft X-ray band, and now only a five times
overabundance of Fe was required. The ionization parameter of the
second component is $\log \xi = 3.397$ resulting in a strong He-like
\fe\ line at 6.7~\kev\ and an inner radius of 5.4~$r_g$. The other
reflector remained relatively neutral with $\log \xi =2.09$ and
strongly blurred ($r_{\mathrm{in}}=1.24$~$r_g$). The photon-index
flattened slightly to $\Gamma=2.41$. While this fit is almost
acceptable there is still a line-like residual above 6~\kev. Adding a
Gaussian component to account for this feature significantly improves
the fit ($\chi^2$/d.o.f.=292/302). This new line has an energy of
6.45~\kev\ and a width of 0.358~\kev, demonstrating that the line
arises from weakly ionized Fe in the accretion disc.

\begin{figure}
\includegraphics[width=0.3\textwidth,angle=-90]{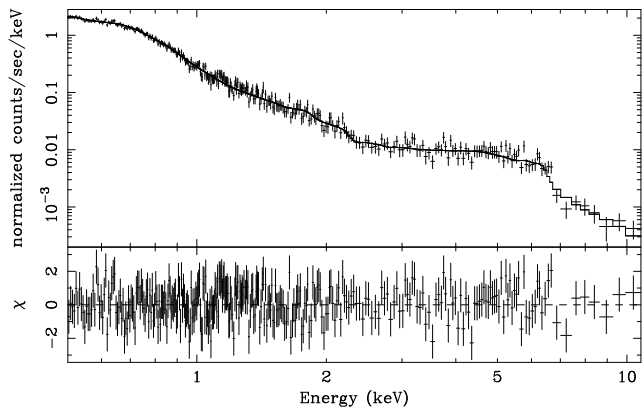}
\caption{The upper panel shows the folded three-reflector model and
EPIC-pn data of \h\ between 0.5 and 11~\kev. The lower panel displays
the residuals to the fit in units of standard deviations.}
\label{fig:1hfit}
\end{figure}

\begin{figure}
\includegraphics[width=0.34\textwidth,angle=-90]{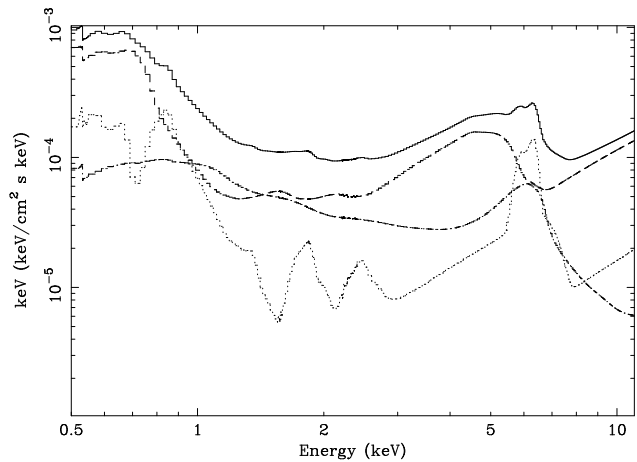}
\caption{The three-reflector model that was successfully fit to \h. The
solid line shows the total model, the dashed line plots the highly blurred
but neutral inner reflector, the dot-dashed line shows the ionized
reflector with $r_{\mathrm{in}}=4.81$~$r_g$, and the dotted line plots
the outer neutral reflector with $r_{\mathrm{in}}=19.8$~$r_g$. These
models do not contain any contribution from the incident power-law with
$\Gamma=2.35$.}
\label{fig:1hmod}
\end{figure}

The two-reflector plus Gaussian emission line model gives a very good
fit to the EPIC-pn data between 0.5 and 11~\kev, however it is
interesting to consider replacing the line component with a third
reflection spectrum.  This model also results in an excellent fit to
the data ($\chi^2$/d.o.f.=287/303; Figure~\ref{fig:1hfit}), but
requires a 7 times solar abundance
of Fe. The three components all have different ionization parameters:
$\log \xi = 1.986$ (with $r_{\mathrm{in}}=2.08$~$r_g$ and
$r_{\mathrm{out}}=7.25$~$r_g$), 3.847 (with
$r_{\mathrm{in}}=4.18$~$r_g$), and 1.722 (with
$r_{\mathrm{in}}=19.8$~$r_g$). The last two reflectors had their outer
radii frozen at 50~$r_g$. The photon index and the inclination angle
for all three components were fixed to have the same values. The best
fit values were $\Gamma=2.35$ and 18.2 degrees, respectively. The
model is shown in Figure~\ref{fig:1hmod}.

\section{Discussion}
\label{sect:discuss}

We have shown that the X-ray spectrum of \h\ is well fitted by a
reflection-dominated spectrum. The incident power-law is mildly steep
with a photon index of 2.35. Furthermore, the model can completely
account for the strong soft excess in this object by a combination of
blurred emission lines and thermal emission from the irradiated
blobs. The geometry envisaged for the source consists of a set of deep
dense sheets or rings, orbiting the central black hole, between which
hard X-rays are emitted. The fact that at least two reflectors are
needed to provide a good fit to the data is not inconsistent with the
model. There should be numerous rings or sheets in the centre of the
disk with (as in the standard coronal picture) multiple hard X-ray
sources active at any one time.  The system is observed at sufficient
inclination that the hard X-ray sources are not directly visible and
only X-rays reflected from the sides of the dense matter are seen.

In fact, the model we have presented is a simplified picture of an 
intrinsically complex situation. In this respect, the interpretation of the 
fit parameters should be consider with some caution. In a more realistic 
situation, the reflection features will be broadened due to both relativistic 
effects and disc turbulence, so that parameters like $r_{\mathrm{in}}$ and 
$r_{\mathrm{out}}$ should not be taken at face value, but 
may be interpreted as the signature of the presence of different 
reprocessors at different locations. Also, a range of ionization parameters 
is to be expected from the complex configuration we envisage, as a
reflector would likely have an ionization gradient on its surface.
 
A consistent model for both the photon index of the input power-law
and the reflection component is obtained if the opening angle of the
gaps between the sheets or rings is less than 40 degrees and the whole
disk inclination is about 20 degrees (Fig.~\ref{fig:gamma-th} and
Sect.~\ref{sect:1h}). The metallicity is still required to be high
(about 5--7 times the Solar value) but less than for other models.  It
is possible that the metallicity can be lowered if multiple reflection
is included (this will have an approximately similar effect to
squaring the reflection spectrum).

The spectral-independent variability of \h\ \citep{bol01} would be
mostly due to changes in the brightness of the hard X-ray source
causing the observed reflection components to vary. Additional
variability may also arise from changes in the geometry of the system,
with sheets or rings obscuring each other as they orbit the black
hole. The geometry may also account for larger scale variations. When
observed by \textit{XMM-Newton}, \h\ was about ten times fainter than
in previous observations \citep{bol01}. This could be due to slight
changes in the geometry and not necessarily a large change in
intrinsic luminosity. If the hard X-ray source was previously more
directly visible (say the opening angle of the sheets or rings was
larger, or the hard X-ray emitting corona extended nearer to the disk
surface) then the observed flux would have been larger. Indeed, the
spectrum observed by \textit{ASCA} when the flux was higher
($\Gamma\simeq 2.27$, \citealt{lei99}), is consistent with being the
unobscured continuum emission peeking through the cold sheets.

It is possible that the model we have developed here for \h\ has a
wider relevance to Seyfert 1 galaxies. The reflection fraction can
exceed unity if the disk is in the form of deep rings or sheets, and
leads to an intrinsic hard power-law steeper than $\Gamma=2$.  \mcg\ has
sometimes been observed with a spectral index of $2.1$ or steeper
\citep*{ve01,sfi01} and its strong iron line suffers far less
variability than its continuum \citep{ve01}. This could be due to
corrugations in its inner disk which may cause the reflection
components to be stronger and more long-lasting than if the accretion
disc was perfectly flat. 

\section*{Acknowledgments}
ACF thanks the Royal Society for support.  DRB acknowledges financial
support from the Commonwealth Scholarship and Fellowship Plan and the
Natural Sciences and Engineering Research Council of Canada. Based on
observations obtained with \textit{XMM-Newton}, an ESA science mission
with instruments and contributions directly funded by ESA Member
States and the USA (NASA).


\bsp 

\label{lastpage}

\end{document}